\documentclass[twocolumn]{article}
\usepackage[cm]{fullpage}
\usepackage[dvips]{graphicx}
\usepackage[fleqn]{amsmath}
\usepackage{amssymb}
\usepackage{latexsym}

\title{Economical ($k,m$)-threshold controlled quantum teleportation}
\author{
Akira SaiToh\footnote{saitoh@alice.math.kindai.ac.jp}$~{}^1$,
Robabeh Rahimi\footnote{rahimi@alice.math.kindai.ac.jp}$~{}^1$, and
Mikio Nakahara\footnote{nakahara@math.kindai.ac.jp}$~{}^{1,2}$\\
${}^1$Research Center for Quantum Computing,
Interdisciplinary Graduate School of\\ Science and Engineering,
Kinki University, 3-4-1 Kowakae, Higashi-Osaka, Osaka 577-8502, Japan\\
${}^2$Department of Physics, Kinki University, 3-4-1 Kowakae,
Higashi-Osaka, Osaka 577-8502, Japan
}
\date{Last modified: 14 June 2009}
\begin{document}
\maketitle
\begin{abstract}

We study a $(k,m)$-threshold controlling scheme for controlled quantum
teleportation. A standard polynomial coding over ${\rm GF}(p)$ with
prime $p>m-1$ needs to distribute a $d$-dimensional qudit with $d\ge p$
to each controller for this purpose. We propose a scheme using $m$ qubits
(two-dimensional qudits) for the controllers' portion, following
a discussion on the benefit of a quantum control in comparison to a
classical control of a quantum teleportation.
~~
\end{abstract}
\section{Introduction}\label{sec1}
Quantum teleportation \cite{B93} has been one of the leading
discoveries followed by numerous quantum information processing
technologies \cite{G99,NC2000}.
Controlled quantum teleportation \cite{Z00,A03,Y04,D05,M07}
is a variant in which a teleportation of a quantum state is
performed under the supervision of controllers.
Schemes using qubits as keys distributed among controllers
\cite{Z00,A03,Y04,D05,M07} have been extensively studied and
economization of required resources has been accomplished
with respect to the number of qubits involved in an entangled
qubit chain used in a scheme. The studies also include
a security discussion on players' cheating controllers \cite{KM06}.

It is expected that a multifunctional quantum network is realized
for a consumer market in future, presumably based on optical fibers.
So far, simple quantum cryptosystems \cite{B92,LMC05} are highly developed
\cite{M95,M96,H02,S02,MLK06,D07} toward the consumer use, which are
mainly used to generate classical shared cryptographic keys.
An advanced quantum network should be used not only for generating a
classical key but for exchanging quantum states. A network with
Einstein-Podolsky-Rosen (EPR) pairs as links \cite{BVK98} is a
plausible form for this purpose;
optical quantum teleportation has already become a well-established
subject supported by many experimental demonstrations reported in,
e.g., Refs.\ \cite{B97,B98,M03}.
Considering an application for office networks, it is demanded that
data transfers can be under the control of multiple supervisors.
Controlled teleportation schemes have been developed to enable this
feature. As the number of usable qubits and the quality of entanglement
enhances, this function is approaching reality.
A function like a vote to permit a transfer is also considered to
be of public demand.

Qubit keys are commonly used in controlled quantum teleportation schemes
for the basic control such that approvals of all the controllers are
required to let the players transfer a quantum state.
In such a control, use of qubit keys is sufficient and extending
each dimension of a quantum system for a key causes a redundancy.
Qudits ($d$-dimensional quantum systems) with $d>2$ have been, in contrast,
considered to be useful as keys when a vote for decision is in order.

It is easy to notice that such functions are enabled by using a
threshold scheme \cite{B79,S79,KGH83,K84,C99}: controlling a
transfer of a quantum state with a certain threshold in the number of
controllers is implemented by secret sharing schemes (see, e.g., a
discussion in Ref.\ \cite{Z05}). This categorizes
controlled quantum teleportation as a combination of quantum secret
sharing and quantum teleportation. A well-known polynomial
coding \cite{S79,C99} over ${\rm GF}(p)$ with prime $p$ larger than the
number $m$ of participants (controllers in our context) is a quick
solution\footnote{Note that $p$ can be equal to $m$ if $m$ is
prime in the context of controlled quantum teleportation
(See Sec.\ \ref{sec3}).}. 
A classical or quantum $(k,m)$-threshold scheme applied to a
shared state of players and controllers enables a control such that a
receiver can recover an original state when and only when $k$ or more
controllers among $m$ provide their keys. There is a variant of quantum
polynomial coding robust against a certain number of cheaters among
participants \cite{M05}.

A drawback to introduce a polynomial coding scheme in controlled quantum
teleportation is that computation is conducted over the field ${\rm
GF}(p)$. Thus each qudit distributed among controllers should have
the dimension $d\ge p$. This also complicates a quantum circuit to
compute a matrix equation for the coding.
For a threshold scheme, i.e., a secret sharing without imposing
access structure\footnote{
See, e.g., Ref.\ \cite{I06} and citations therein for secret sharing schemes
involving access structures.}, the dimension of each key (share) using
best known classical protocols \cite{KT06,K08} is $O(2^m)$ (namely,
$O(m)$ bits for each key) for a long secret. These protocols require the
bit length of a secret $O(m)$ and that of a key at least as much as the
length of a secret. For a short secret, a best known classical protocol
is Shamir's one \cite{S79} in which each key has the dimension $\ge p>m$.
It is thus not motivating to simply make a quantum extension
of the classical protocols.

It was reported that an $(m,m)$-threshold secret sharing of a classical
secret is achieved by a sort of key distribution without entanglement
\cite{GG03} and an $(m,m)$-threshold controlled quantum teleportation is
achieved by using classical keys \cite{ZM04}. The latter one is easily
extended to a $(k,m)$-threshold controlling scheme. Nevertheless,
schemes using qudits for controllers' portion are more secure than
the one using only classical keys as we will discuss in Sec.\ \ref{secCCQT}.
There is a recently-proposed graph-state formalism \cite{MS08}
to produce quantum secret sharing systems using only qubits.
This approach has achieved the systems for several particular $(k,m)$
and not for general $(k,m)$.
We will construct a scheme which is not a direct extension of these
approaches.

In this paper, we introduce a $(k,m)$-threshold controlled quantum
teleportation scheme over ${\rm GF}(p)$ with prime $p>m-1$ using
$m$ qubits for the controllers' portion in addition to classical
information distributed to the controllers. This is achieved by
a hybrid of classical and quantum protocols.
It enables a reduction in the dimension of each qudit to two for general
$(k,m)$, which is a significant improvement for realizing the scheme.
In contrast to classical systems, the dimension of a quantum system is
limited and usually a large-dimensional qudit is implemented by multiple
qubits. The size of a qudit register is also very limited in the present
technology. The number of available qubits is, so far, twelve or
less \cite{N06}.

To begin with, a standard controlled quantum teleportation scheme
is briefly described in Sec.\ \ref{sec2}. On the basis of the scheme, 
a $(k,m)$-threshold controlled quantum teleportation using a
polynomial coding is briefly explained in Sec.\ \ref{sec3}.
A reduction in required resources for the threshold-control scheme is
accomplished in Sec.\ \ref{sec4}: First, a classical control
of a quantum teleportation is discussed in Sec.\ \ref{secCCQT}.
Second, a control of a quantum teleportation using $m$
qubits for the controllers' portion is introduced in Sec.\ \ref{secEQCQT}.
Third, an economization of Bob's operations is accomplished
in Sec.\ \ref{secEBO}. An explicit preparation of an initial state for
this economized scheme is shown in Sec.\ \ref{sec5} together with
the operational complexity of the whole process. We discuss
advantages and disadvantages of the scheme in Sec.\ \ref{sec6}.
Section\ \ref{sec7} summarizes our results.

\section{A Standard Controlled Quantum Teleportation}\label{sec2}
\begin{figure}[hpbt]
\begin{center}
 \scalebox{0.6}{\includegraphics{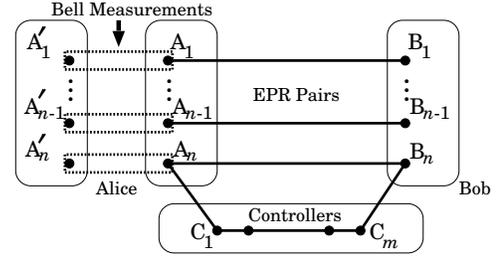}}
\caption{\label{fig1}
Illustration of qudit sharing for a controlled quantum teleportation.}
\end{center}
\end{figure}

Consider a controlled quantum teleportation using ($n-1$) EPR pairs
shared by Alice and Bob, and a single quantum system shared by Alice,
Bob, and $m$ controllers. Alice tries to send an $n$-qubit state
\[
\sum_{x_1\ldots x_n=0\ldots 0}^{1\ldots1}
p_{x_1\ldots x_n}|x_1\ldots x_n\rangle_{{\rm A'}_1...{\rm A'}_n}
\]
of the system ${\rm A'}$ to Bob. A standard quantum teleportation protocol
works fine for the ($n-1$)-EPR-pair channel consisting of ($n-1$) pairs
${\rm A}_1{\rm B}_1,...,{\rm A}_{n-1}{\rm B}_{n-1}$. The remaining channel,
${\rm A}_n{\rm B}_n$, is under the control of $m$ controllers
${\rm C}_1,...,{\rm C}_m$.
The setup of the quantum system is illustrated in Fig.\
\ref{fig1}. Here it should be noted that, although we consider
the control of a single channel here, it is straightforward to attach
controllers to each channel. Thus let us limit setups to the illustrated
one in the following.
The initial state of the illustrated system is given by
\[\begin{split}
|{\rm A'ABC}\rangle=&\left(\sum_{x_1\ldots x_n=0\ldots 0}
^{1\ldots1} p_{x_1\ldots x_n}|x_1\ldots x_n\rangle_{{\rm A'}_1...{\rm A'}_n}\right)\\
&\otimes
 \left(\bigotimes_{l=1}^{n-1}\frac{1}{\sqrt{2}}\sum_{y_{{}_l}=0}^1
|y_{{}_l}\rangle_{{\rm A}_{{}_l}}|y_{{}_l}\rangle_{{\rm B}_{{}_l}}\right)\\
&\otimes
|\xi\rangle_{\rm A_{\it n}B_{\it n}C_1...C_{\it m}}
\end{split}\]
with
\begin{equation}\label{eqkappa}
|\xi\rangle_{\rm A_{\it n}B_{\it n}C_1...C_{\it m}}
=\frac{1}{\sqrt{2}}\sum_{y=0}^1|yy\rangle_{\rm A_{\it n}B_{\it n}}
|\kappa(y)\rangle_{{\rm C}_1...{\rm C}_m}
\end{equation}
a shared state involving the controllers' portion
$|\kappa(y)\rangle_{{\rm C}_1...{\rm C}_m}$
to be engineered for a tailored controlling scheme.

Let us recall the well-known relation
\[
 |x,y\rangle=\frac{1}{\sqrt{2}}\sum_{i=0}^1 (-1)^{x\cdot i}
|B_{i,x\oplus y}\rangle,
\]
where $|x,y\rangle$ is a computational basis vector and
$|B_{i,j}\rangle = (1/\sqrt{2})\sum_{x=0}^1(-1)^{x\cdot i}|x, j\oplus x\rangle$
is the $(i,j)$th Bell basis vector (here, $j=x\oplus y$,
i.e., $y=j\oplus x$).
With this relation, we can rewrite the initial state as
\begin{equation}\label{eqrewritten}
\begin{split}
&|{\rm A'ABC}\rangle\\
&=\frac{1}{2^n}\sum_{x_1...x_{n}}p_{x_1...x_{n}}
\biggl[\biggl(\bigotimes_{l=1}^{n-1}
\sum_{i_l=0}^1 \sum_{j_{{}_l}=0}^1 (-1)^{x_{{}_l}\cdot i_{{}_l}}\\
&\times
|B_{i_l,j_l}\rangle_{{\rm A'}_l{\rm A}_l}
|j_{{}_l}\oplus x_{{}_l}\rangle_{{\rm B}_l}\biggr)
\otimes\biggl(
\sum_{i_{n}=0}^1 \sum_{j_{n}=0}^1 (-1)^{x_{n}\cdot i_{n}}\\
&\times
|B_{i_{n},j_{n}}\rangle_{\rm {A'}_{\it n}A_{\it n}}
|j_{n}\oplus x_{n}\rangle_{\rm B_{\it n}}
|\kappa(j_{n}\oplus x_{n})\rangle_{\rm C_1...C_{\it m}}\biggr)\biggr].
\end{split}
\end{equation}

As is usual for a standard quantum teleportation,
Alice makes Bell measurement on each $({\rm A'}_l{\rm A}_l)$
pair and obtains an outcome $(i_{{}_l},j_{{}_l})$.

Bob receives information $\{(i_{{}_l},j_{{}_l})\}$ from Alice and applies
$\bigotimes_{l=1}^{n-1}Z^{i_l}X^{j_l}$ to ${\rm B}_1...{\rm B}_{n-1}$.
This changes the state of each ${\rm B}_l$ (of each term in the summation)
from $(-1)^{x_{{}_l}\cdot i_{{}_l}}|j_{{}_l}\oplus x_{{}_l}\rangle_{{\rm B}_l}$
to $|x_{{}_l}\rangle_{{\rm B}_l}$ and hence the teleportation process for
the original state of ${\rm A}'$ is completed up to $l = n-1$.

To complete the recovery of the original state of ${\rm A}'$ at
Bob's side, he has to apply a certain operation to ${\rm B}_n$.
This requires Bob to know the effect of controllers' measurements on
$|\kappa(j_{n}\oplus x_{n})\rangle_{\rm C_1...C_{\it m}}$.
Operations Bob has to apply to ${\rm B}_n$ depend on the controlling
scheme.

A popular controlled quantum teleportation scheme is the case where
$|\kappa(y)\rangle_{\rm C_1...C_{\it m}}$ is set to
$|y\ldots y\rangle_{\rm C_1...C_{\it m}}$, i.e.,
$|\xi\rangle_{{\rm A}_n{\rm B}_n{\rm C}_1...{\rm C}_m}$ is set to a
Greenberger-Horne-Zeilinger (GHZ) state, and each controller makes
a measurement in the $X$ basis. Measuring ${\rm C}_s$ in the $X$
basis results in the phase factor
$(-1)^{y\cdot(h_s==-1)}$ depending on the outcome $h_s$ (here,
$s\in \{1,...,m\}$; the operation ``$==$'' returns one if its two
arguments are equal and zero otherwise)\footnote{
This is because a computational basis state $|c\rangle$ of a qubit can be
written as
$
|c\rangle=\frac{1}{\sqrt{2}}\sum_{h=+1,-1}(-1)^{c\cdot(h==-1)}|h\rangle
$ using the $X$-basis states $|h\rangle$.}.
In this case, Bob first applies $Z^{i_n}X^{j_n}$ to
${\rm B}_n$. In addition, he applies a single $Z$ gate to ${\rm B}_n$
for recovery if the number of $-$'s in the controllers' outcomes
($\in\pm$) is odd.
The final state of Bob after these operations becomes
$\sum_{x_1...x_{n}}p_{x_1...x_{n}}
\bigotimes_{l=1}^{n}|x_{{}_l}\rangle_{{\rm B}_l}$. The
teleportation is successful in this way.

In this contribution, we aim to introduce a threshold-control scheme in
which the shared state $|\xi\rangle_{\rm A_{\it n}B_{\it n}C_{1}...C_{\it m}}$
is engineered to be different from the GHZ state. We continue to
concentrate on the case where only a single qubit of Bob is
under the control, as illustrated in Fig.\ \ref{fig1}; this is because it is
straightforward to extend the scheme so that multiple qubits are under
the control. Such an extension will be considered only in Sec.\ \ref{secEBO}.

\section{$(k,m)$-Threshold Controlled Quantum Teleportation}\label{sec3}
The functionality of threshold control is achieved by
engineering the initial setup of the shared state given
by Eq.\ (\ref{eqkappa}). Let us begin with a rather expensive scheme
that is a straightforward extension of a classical secret sharing.
In this scheme, the state of the controllers' portion is assumed
to be in the following form
\begin{equation}\begin{split}\label{eqqss}
&|\kappa(y)\rangle_{{\rm C}_1...{\rm C}_m}\\
&=\frac{1}{\sqrt{\#\mathcal{S}}}
\sum_{c_1...c_m\in \mathcal{S}}
e^{iy\theta(c_1...c_m)}|c_1...c_m\rangle_{{\rm C}_1...{\rm C}_m},
\end{split}\end{equation}
where $y=j_{n}\oplus x_{n}$, $c_s\in\{0,\ldots,d-1\}$
($s\in\{1,...,m\}$ and $d$ is a positive integer), and
$\mathcal{S}$ is a certain set of $m$-digit strings;
the computational basis for each ${\rm C}_s$ is chosen
arbitrarily and known by the $s$th controller.
There are two conditions for the state appropriate for a
$(k,m)$-threshold scheme:\\
(i) $\theta(c_1...c_m)$ cannot be uniquely determined
unless the sequence $c_1...c_m$ is completely specified
(i.e., unless the variables $c_1,...,c_m$ are all specified).\\
(ii) A string $c_1...c_m$ is uniquely determined in $\mathcal{S}$
by fixing any $k$ digits of $c_1,...,c_m$.\\
Under these conditions, $k$ controllers' measurements result in
a single surviving vector $|c_1...c_m\rangle$. Thus Bob can
recover the original state by the following process (in addition to the
usual process for the original quantum teleportation for $l=1,...,n-1$).
First Bob changes the phase factor $e^{iy\theta(c_1\ldots c_m)}$ to unity
by applying
\[
 \begin{pmatrix}1&0\\0&e^{-i\theta(c_1\ldots c_m)}\end{pmatrix}
\]
to the qubit ${\rm B}_n$. Second Bob applies $Z^{i_n}X^{j_n}$ to
${\rm B}_n$ as is usual for a quantum teleportation.
In these steps, Alice's Bell measurement on ${\rm A'}_n{\rm A}_n$
prior to controllers' measurements does not affect Bob's operation
to recover the phase appearing as a result of controllers' measurements.
This is clear from the initial state described in
Eq.\ (\ref{eqrewritten}); the parameter $y=j_{n}\oplus x_{n}$ is common
in the states of ${\rm B}_n$ and controllers' portion. 

Let us turn to a system construction for satisfying the conditions.
Condition (i) is satisfied by setting
$\theta(c_1...c_m)=\sum_{s=1}^m c_s\pi/d$.
Such $\theta(c_1...c_m)$ cannot be uniquely determined
unless the variables $c_1,...,c_m$ are completely specified.
A quantum circuit to attach the phase factors is easily realized:
from $s=1$ to $m$, applying an ${\rm A}_n$-controlled ${\rm
C}_s$-controlled phase gate with the phase $c_s \pi/d$, acting on
a work qubit accomplishes this task.
For the condition (ii), we will find a proper set $\mathcal{S}$ of
$m$-digit vectors $c_1...c_m$ by a certain coding scheme.
This is the main concern as the resource for quantum information
processing is limited in the current technologies; a coding with
a small resource is desirable.

We revisit the theory of Karnin {\em et al.} \cite{KGH83},
which describes a sufficient condition for coding in a threshold scheme,
and evaluate a polynomial-coding scheme \cite{S79,C99} as a typical example.

We begin with a well-known property of a linear matrix equation.
\begin{quote}{\em Proposition:} 
A solvable matrix equation over ${\rm GF}(p)$,
\[
 A{\bf x}={\bf b}
\]
with $p$ a prime number, $A\in {\rm GF}(p)^{f\times g}$,
${\bf x}\in {\rm GF}(p)^g$, and
${\bf b}\in {\rm GF}(p)^f$ has a unique solution ${\bf x}$ if
and only if $f\ge g$ and $A$ has a full rank.
\end{quote}
The proof is similar to the case of a real number field
(see, e.g., Ref.\ \cite[pages 96,103]{LT85}).\\
{\em Proof|}
(i) First we prove that the solution ${\bf x}$ is unique if
$f\ge g$ and $A$ has a full rank.
Assume that there are two solutions ${\bf x}_1$ and ${\bf x}_2$.
Then, $A{\bf x}_1=A{\bf x}_2$.
Let us pick up $g$ rows of $A$ appropriately to generate $\tilde A$
so that $\tilde A$ has a full rank. This is possible because
otherwise the number of linearly independent row vectors of $A$ should be
less than $g$, which is a contradiction. We have thus
$\tilde A {\bf x}_1 = \tilde A {\bf x}_2$ with the square full-rank matrix
$\tilde A$. The matrix $\tilde A$ can be reduced to a diagonal matrix
with nonzero diagonal elements by basic operations; thus
${\rm det}\tilde A\not = 0$.
Consequently, ${\bf x}_1={\bf x}_2$ because ${\tilde A}^{-1}$ exists.
This is a contradiction.
(ii) Second we prove that $f\ge g$ and $A$ has a full rank if ${\bf x}$
is a unique solution. The contraposition of this statement is that
solution ${\bf x}$ is not unique if $f<g$ or ${\rm rank}~A$ is
less than ${\rm min}(f,g)$.
This is easily shown to be true.$~\square$

On the basis of this proposition, a coding of our interest is achieved
by a matrix equation for $(f,g)=(m,k)$ with a full-rank matrix $A$
such that striking any $(m-k)$ rows keeps the rank full \cite{KGH83}.
Given a matrix equation
\begin{equation}\label{eqAxc}
 A {\bf x}= (c_1,\ldots,c_m)^t
\end{equation}
with such $A$ which is notified to Bob, we prepare
$|\kappa(y)\rangle_{{\rm C}_1\ldots{\rm C}_m}$ as
\[\begin{split}
&|\kappa(y)\rangle_{{\rm C}_1\ldots{\rm C}_m}\\
&=\frac{1}{\sqrt{\# \mathcal{S}_{\bf x}}}
\sum_{c_1\ldots c_m\in\mathcal{S}_{\bf x}}
e^{iy\theta(c_1\ldots c_m)}|c_1\ldots c_m\rangle_{{\rm C}_1\ldots{\rm C}_m}
\end{split}\]
where $\mathcal{S}_{\bf x}$ is a set of $c_1\ldots c_m$
corresponding to ${\bf x}\in{\rm GF}(p)^k$ (hence
$\# \mathcal{S}_{\bf x}\le p^k$).
Suppose that at least $k$ controllers measure their qudits
in the computational basis. Then ${\bf x}$ is fixed uniquely because the
matrix equation becomes solvable by using the rows corresponding to
the fixed $c_s$'s. This implies that the superposition is then resolved.
Bob can determine the phase factor that he should modify by receiving
at least $k$ outcomes from controllers.

We have seen a common construction of a threshold scheme.
Required resources for the threshold scheme are mostly dependent
on the choice of the matrix $A$. There are two practical ways
among many \cite{KGH83,K84}.
One is the matrix in the form
\[
 A_{\rm (i)}=\left(~I~|~T~\right)^t
\]
with $I$ the $k\times k$ identity matrix and $T$ a $k \times (m-k)$
strictly totally positive matrix \cite{K68,A87}.
Any minor of $T$ is nonzero positive from the
definition of strict total positivity; hence any $k$ rows of $A$
build up a square matrix with nonzero determinant. Thus a matrix
equation with $A_{\rm (i)}$ can be used for the threshold scheme.
A drawback is the difficulty to find a strictly totally positive
matrix $T$ for sufficiently small prime $p$. A known systematic
construction \cite{C98} for strictly totally positive matrices
uses the largest element of $T$ growing exponentially in
${\rm dim}~T$. Thus $p$ also grows exponentially if we follow the
construction. A manual optimization is indispensable.

The other is a Vandermonde matrix used in the
well-known Shamir's scheme \cite{S79},
\begin{equation}\label{Vmatrix}
 A_{\rm (ii)}= V_{m,k}=
\begin{pmatrix}
1&x_1&x_1^2&\cdots&x_1^{k-1}\\
1&x_2&x_2^2&\cdots&x_2^{k-1}\\
\vdots&\vdots&\vdots&\vdots&\vdots\\
1&x_m&x_m^2&\cdots&x_m^{k-1}
\end{pmatrix}~~~{\rm mod}~p
\end{equation}
with mutually different $x_i$'s with prime $p>m-1$. 
($p>m-1$ is necessary to set $x_i$'s mutually different.)
Striking $(m-k)$ rows generates a square Vandermonde matrix
and it is non-singular when $x_i$'s are mutually different (see, e.g.,
pages 43 and 219 of Ref.\ \cite{PS76}).
Hence a matrix equation with $A_{\rm (ii)}$ can be used for
the threshold scheme. This matrix has been known to be economical
because $p$ increases linearly in $m$. Each qudit distributed to
a controller should have the dimension $d\ge p$, consequently.
Nevertheless, one may need to further reduce the dimension considering
the poor resources of presently available qudit systems \cite{N06,Na06}.

\section{Economizing the Threshold Control Scheme}\label{sec4}
The dimension of each digit distributed to a party (a controller in the
present context) is often evaluated by using the scale of ``bit length''
in conventional secret sharing schemes. Each key is $O(\log m)$-bit
long \cite{S79} ($O(m)$-bit long \cite{KT06,K08}) in the best known
protocols for a short secret with the bit length $O(\log m)$
(for a long secret with the bit length $O(m)$) although
this has not been taken as a drawback at all since classical bits are
very cheap. Nevertheless, in quantum protocols one should not consume
many qubits for individual quantum systems.
It is of our concern to find a smaller dimension for each
controller's qudit facing a limited resource of a quantum system.

One way is to abandon the use of quantum systems for controllers'
portion and instead use a classical threshold scheme to control a quantum
teleportation.
A controlled teleportation proposed by Zhang and Man \cite{ZM04}, in the
context of $(m,m)$ threshold, uses classical keys shared by Alice and
controllers for encoding Alice's messages, which can be easily extended
to a general threshold-control scheme.
Here, we introduce a different scheme where controllers' qudits are
simply replaced by classical digits. We will face the fact that
classical control schemes are indeed economical but their security is
based on classical keys. Using qudits is found to be more robust
against Bob's physical-access attack.

Our interest is to find such a robust scheme with a simple quantum state
for the controllers' portion. It is shown to be constructed by
using the matrix equation with the Vandermonde matrix (\ref{Vmatrix})
and qubit states distributed to the controllers. We further perform an
economization in the number of Bob's operations, which is useful for an
extension in which multiple EPR channels are under controllers' control.

\subsection{Classically-Controlled Quantum Teleportation}\label{secCCQT}
As we mentioned, the simplest way of economization is to use classical
control digits instead of quantum ones. This is easily achieved by setting the
state (not a state, actually) of the controllers' portion to be a scalar 
\[
 |\kappa(y)\rangle_{{\rm C}_1...{\rm C}_m} = e^{iy\theta(c_1,...,c_m)}
\]
with phase $\theta$ dependent on integers $c_1,...,c_m$, the classical
keys of a certain classical threshold scheme. Bob can modify this
scalar factor by applying ${\rm diag}[1, e^{-i\theta(c_1,...,c_m)}]$ to
his $n$th qubit ${\rm B}_n$ if he can gather at least $k$ of the keys. 

The security of the scheme is dependent on the classical scheme.
Indeed, classical keys can be securely distributed by
using a quantum key distribution (see Ref.\ \cite{G02} and references
therein) and the risk of an interception during the key distribution
is negligible. Classical keys are, however, easily copied by careless
controllers. A possible drawback of the scheme is that
controllers cannot stop Bob from recovering Alice's original state
if Bob manages to obtain at least $k$ of the keys without consent of the
controllers.
In contrast, in a threshold scheme using qudits, the operations for a
recovery of the original state are unfixed until $k$ controllers make
measurements. This fact makes the quantum one more robust: Recall
that the computational basis for ${\rm C}_s$ in Eq.\ (\ref{eqqss}) is
not necessarily known in public.
The correct basis for a measurement can be left unknown to Bob. Then,
Bob cannot obtain $c_s$ by a physical access to ${\rm C}_s$ unless he
also has an access to the information on the measurement basis. Thus
the cost for Bob to steal $c_s$'s in the scheme is more than that in
the classically controlled one. This logic is similar to
the one utilized for keeping the security of dealer-player
communications in some quantum threshold schemes \cite{GG03,MS08}. 

It should be mentioned that robustness depends on the type of protocol
violations. Let us discuss a different type of violation that can be made
by Bob. It is a common occasion that controllers do not want Bob to process
further with a teleported state before they officially vote for their
decision. A violation in this regard occurs when $k$ or more
controllers are friendly to Bob and they measure their qudits or digits
and send the outcomes before the voting starts officially.
This violation in the schedule of the procedure cannot be prevented
even if qudits are distributed.

In the following, an economical quantum control scheme is introduced
in order to reduce the resource for controllers' quantum system as we
have mentioned. It is now our additional motivation to resolve the
schedule violation problem due to friendly controllers.
\subsection{Economical Quantum Threshold-Controlled Quantum Teleportation}\label{secEQCQT}
As is discussed above, a quantum threshold scheme has a classically
unachievable property, namely  that the operations for a recovery of the
original state are unfixed until controllers make measurements and the
measurement bases can be left unnotified to Bob. As we
have mentioned, our aim is to achieve such a scheme using
small-dimensional systems for controllers' portions. Here, we propose a
quantum protocol for the $(k,m)$-threshold controlled quantum
teleportation with $m$ qubits distributed to the controllers. It also
resolves the problem of the possible violation in the voting schedule.

It is implemented with the following state for the controllers' portion:
\begin{equation}\label{econkappa}
\begin{split}
&|\kappa(y)\rangle_{{\rm C}_1...{\rm C}_m}\\
&=\frac{1}{\sqrt{2^m}}
\bigotimes_{s=1}^m [
  e^{i2\pi y c_s/p}|{\tilde 0}_s\rangle
+ e^{i2\pi \neg{y} c_s/p}|{\tilde 1}_s\rangle]_{{\rm C}_s},
\end{split}
\end{equation}
where $|{\tilde 0}_s\rangle$ and $|{\tilde 1}_s\rangle$ are the basis
vectors of the $s$th controller's chosen basis; $\neg{y}$ is a logical
negation of $y$ and $c_s$'s are the keys of the following common
classical polynomial-coding threshold scheme.
The keys are generated from Eq.\ (\ref{eqAxc}) using a Vandermonde matrix for
$A$ and a certain fixed vector for ${\bf x}$ [all the matrix and vector
elements are in ${\rm GF}(p)$]. The matrix $A$ is notified to Bob
and ${\bf x}$ is hidden. Thus $k$ or more keys are required for Bob
to determine the remaining keys from Eq.\ (\ref{eqAxc}).

The protocol imposed to controllers is as follows.\\
(I) Each controller ${\rm C}_{s,{\rm agree}}$
who agrees to allow Bob to recover Alice's original
state measures her/his qubit in the basis
$\{|{\tilde 0}_s\rangle,|{\tilde 1}_s\rangle\}$ and sends its outcome
$r_s\in\{0,1\}$ to Bob.\\
(II) ${\rm C}_{s,{\rm agree}}$ also sends her/his key $c_s$ to Bob.\\
(III) Each controller ${\rm C}_{s,{\rm disagree}}$ who disagrees
to allow Bob to recover Alice's original state does not make any action
until she/he receives a contact from Bob.\\
(IV) When a solicit is sent from Bob,
${\rm C}_{s,{\rm disagree}}$ must measure her/his qubit in the basis
$\{|{\tilde 0}_s\rangle,|{\tilde 1}_s\rangle\}$ and send its outcome
$r_s\in\{0,1\}$ to Bob.

The protocol imposed to Bob is as follows.\\
(i) Bob receives $(i_{{}_l},j_{{}_l})$'s from Alice. Bob applies
$\bigotimes_{l=1}^{n-1}Z^{i_l}X^{j_l}$ to ${\rm B}_1...{\rm B}_{n-1}$.\\
(ii) Bob waits for at least $k$ pairs of $(r_s,c_s)$'s
sent from the controllers.\\
(iii) Bob calculates the remaining $c_s$'s by substituting
the obtained $c_s$'s into Eq.\ (\ref{eqAxc}) if at least $k$ pairs
are obtained; aborts otherwise.\\
(iv) Bob sends solicits to the controllers who did not
send information to him.\\
{\em Note:} it is possible to count his solicits. Therefore,
he cannot cheat by sending more than ($m-k$) solicits in this stage
even when he succeeds in stealing $k$ or more $c_s$'s beforehand.\\
(v) Bob receives the remaining $r_s$'s.\\
(vi) Bob modifies the phase factor due to controllers' measurements
in his phase recovery process described below.\\
(vii) Bob applies $Z^{i_n}X^{j_n}$ to ${\rm B}_n$.
\paragraph*{Bob's phase recovery process corresponding to the controllers'
measurements|}
The $m$ controllers' measurements (of course, after Alice's Bell
measurements) make the component state for
${\rm A'}_n{\rm A}_n{\rm B}_n{\rm C}_1...{\rm C}_m$ in Eq.\
(\ref{eqrewritten}), with the controllers' state (\ref{econkappa}) in
the present context, evolve into the state
\[
\sum_{y=j_n\oplus x_n=0}^1
(-1)^{x_n\cdot i_n} |B_{i_n,j_n}\rangle_{{\rm A'}_n{\rm A}_n}
|y\rangle_{{\rm B}_n}
|\widetilde{\kappa}(y)\rangle_{{\rm C}_1...{\rm C}_m}
\]
with
\[\begin{split}
&|\widetilde{\kappa}(y)\rangle_{{\rm C}_1...{\rm C}_m}\\
&=\frac{1}{\sqrt{2^m}}
\bigotimes_{s=1}^m \exp[i 2\pi (\neg)^{r_{s}}y c_s/p]
|r_{s}\rangle_{{\rm C}_{s}}.
\end{split}\]
The phase factor
\[
\prod_s \exp[i 2\pi (\neg)^{r_{s}}y c_s/p]
\]
should be canceled by Bob before the normal recovery operation
is performed in step (vii).
The cancellation of the factor with probability one is possible if and
only if all the controllers follow the protocol and at least $k$
of them send $c_s$'s to Bob (otherwise he will get an uncertain
state as his operations become a guesswork).
Since $c_s$'s are classical keys of a $(k,m)$ threshold scheme,
$k$ or more of them are necessary and sufficient to find out
all of them. Bob can eliminate the phase by applying
\begin{equation}\label{eqrecov}
\prod_s {\rm diag}[\exp(-i2\pi r_s c_s/p), \exp(-i2\pi\neg r_s c_s/p)]
\end{equation}
to his $n$th qubit ${\rm B}_n$.
This operation is unaffected by Alice's Bell measurement on
${\rm A'}_n{\rm A}_n$ as is clear from the initial state described
in Eq.\ (\ref{eqrewritten}).

In this way, the $(k,m)$-threshold control is realized by using only
qubit systems for individual portions.
One may notice that it is a hybrid scheme in the sense that the
$(k,m)$ threshold is realized by a classical threshold scheme.
Although classical keys are used, we claim that this scheme
possesses a property thanks to a quantum control; the recovery
operations are unfixed unless all the controllers make measurements, and
the measurement bases are unknown to Bob.
The measurements are not completely performed unless there are
at least $k$ ${\rm C}_{s,{\rm agree}}$'s and 
Bob sends at most ($m-k$) solicits to ${\rm C}_{s,{\rm disagree}}$'s.
Hence, in order to cheat under the protocol, Bob needs to collect
$c_s$'s, qubits, and measurement bases of at least $k$ controllers.

The scheme has an advantage over the previously-introduced quantum
scheme in the sense that it is secure against the possible schedule
violation due to friendly controllers. Owing to step (IV), Bob
has to wait for an official voting time to obtain Alice's original state
unless the controllers violate the voting schedule all together.
\subsection{Economization in Bob's Operations}\label{secEBO}
We have shown an economized quantum (or hybrid) controlled quantum
teleportation using the state (\ref{econkappa}) for the controllers'
portion, in which the dimension of each distributed qudit has been
reduced to two. As two is the minimal dimension for a nontrivial
quantum system, it is optimized with respect to the dimension of a
Hilbert space of each portion. Here we consider an economization of
Bob's operations to eliminate the phase factor coming from controllers'
measurements.

\begin{figure}[hpbt]
\begin{center}
\scalebox{0.6}{\includegraphics{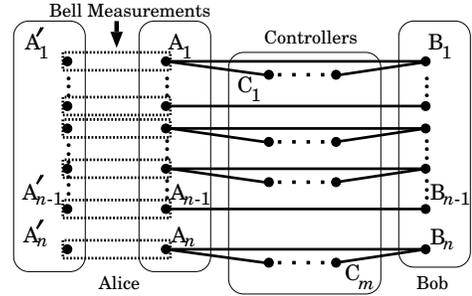}}
\caption{\label{fig2} A straightforward extension of the system setup
shown in Fig.\ \ref{fig1}, such that controllers are assigned to
multiple channels. Each ${\rm A}_l{\rm B}_l$ pair is not necessarily
under the supervision of controllers.}
\end{center}
\end{figure}

The number of Bob's single-qubit operations for the recovery is not
important as far as a single EPR channel is under the controllers' control
because Eq.\ (\ref{eqrecov}) reduces to a single operation.
It is, however, not negligible in case we extend the setup to the one
illustrated in Fig.\ \ref{fig2}, in which we may have multiple shared
states. Each controller is assigned to a single ${\rm A}_l{\rm B}_l$
pair and the total number of controllers is $m$.

Let us use the same state as in Eq.\ (\ref{econkappa}), except the labels of
controllers, for each of the controllers' portions (represented by
dotted lines inside the square of ``Controllers'' in Fig.\ \ref{fig2}).
We change the protocol for controllers in the following way.\\
(A) A controller who Agrees to allow Bob to recover the original state
makes a measurement on her/his qubit in the basis
$\{|{\tilde +}_s\rangle, |{\tilde -}_s\rangle\}$ with
$|{\tilde \pm}_s\rangle = |{\tilde 0}_s\rangle
\pm |{\tilde 1}_s\rangle$.
She/he sends Bob the outcome $u_s\in\{\tilde +, \tilde -\}$ together with
$c_s$.\\
(D) A controller who Disagrees does not make any action. However
when a solicit is sent from Bob, a controller
who Disagrees must make a measurement on her/his qubit in the basis
$\{|{\tilde 0}_s\rangle, |{\tilde 1}_s\rangle\}$ and send Bob the
outcome $v_s\in\{0,1\}$.\\~~

To understand the effect of a measurement, let us decompose each component
of the state (\ref{econkappa}) in the following way:
\[\begin{split}
\frac{1}{\sqrt{2}}
[e^{i2\pi y c_s/p}|{\tilde 0}_s\rangle
+& e^{i2\pi \neg{y} c_s/p}|{\tilde 1}_s\rangle]\\
=\frac{1}{2\sqrt{2}}
\biggl[
&(e^{i 2\pi y c_s/p}+e^{i 2\pi \neg{y} c_s/p})|{\tilde +}_s\rangle\\
&+(e^{i 2\pi y c_s/p}-e^{i 2\pi \neg{y} c_s/p})|{\tilde -}_s\rangle
\biggr].
\end{split}\]

First consider the case (A).\\
(A-i) Suppose that the outcome of a measurement
by the $s$th controller is $\tilde +$. Then, the unnormalized
phase factor owing to this measurement is 
\[
 e^{i2\pi y c_s/p}+e^{i2\pi \neg y c_s/p} = 1+e^{i2\pi c_s/p},
\]
which is a global phase uncorrelated with $y$. Thus Bob does not have to
modify it.\\
(A-ii) Suppose that the outcome of a measurement
by the $s$th controller is $\tilde -$. Then, the unnormalized phase
factor owing to this measurement is 
$
 e^{i2\pi y c_s/p}-e^{i2\pi \neg y c_s/p},
$
which can be regarded as
\[
 \left\{
\begin{array}{ll}
\alpha&(y=0)\\-\alpha&(y=1)
\end{array}
\right.
\]
with $\alpha=1-e^{i2\pi c_s/p}$. Thus Bob can modify the factor
by applying $Z$ to some ${\rm B}_l$ for which the controller's qubit
is originally connected in Fig.\ \ref{fig2}.

Second consider the case (D).\\
The phase factor due to the measurement, with the outcome
$v_s\in\{0,1\}$, is $\exp[i 2\pi (\neg)^{v_{s}}y c_s/p]$.
This can be modified by applying
\[
{\rm diag}[\exp(-i2\pi v_s c_s/p), \exp(-i2\pi \neg v_s c_s/p)]
\]
to a proper ${\rm B}_l$,
which is possible if Bob knows $c_s$, namely, if $k$
or more controllers follow (A).

In addition to the above recovery of the phase factors corresponding to
controllers' measurements, Bob applies $Z^{i_l}X^{j_l}$ to each $B_l$,
as usual, to recover the phase factors corresponding to Alice's Bell
measurements.

Let us estimate the reduction in the number of operations that Bob has to
apply to ${\rm B}_l$'s for eliminating the phase factors due to
controllers' measurements. Let us consider the worst case where
the $m$ controllers are assigned to mutually different channels.
In our previous protocol, the number of the operations is $m$.
In the present protocol, it is, on average, $t/2+(m-t)=m-t/2$ with
$t$ the number of controllers who agree to allow Bob to recover the
original state. (Of course, Bob cannot recover the state when $t<k$.)

\section{Operational Complexity}\label{sec5}
In the previous section, a quantum (or hybrid) $(k,m)$-threshold
controlled quantum teleportation using qubits (without qudits
whose dimension is more than two) has been constructed.
We will count the number of single-qubit operations and that of
two-qubit operations by following the whole process.
Here, we consider the case where the controllers are attached to the
$n$th EPR channel ${\rm A}_n{\rm B}_n$ and regard Bob's recovery
operation corresponding to each controller's measurement as a single
operation, for simplicity. The number of operations is unchanged by
employing the setup illustrated in Fig.\ \ref{fig2}.

The process is the same as the original quantum teleportation \cite{B93}
except for the measurements and operations acting on the shared state
of ${\rm A}_n{\rm B}_n{\rm C}_1...{\rm C}_m$, as we have seen in Sec.\
\ref{sec2}.

For the part without controllers in Fig.\ \ref{fig1}, there are ($n-1$)
EPR pairs prepared between Alice and Bob. The quantum circuit for
preparing the EPR states involves ($n-1$) Hadamard gates and the same
number of controlled-{\small NOT} (CNOT) gates. Alice makes ($n-1$) Bell
measurements between the system ${\rm A'}_1...{\rm A'}_{n-1}$ and the
system ${\rm A}_1...{\rm A}_{n-1}$ and sends the outcomes
($i_{{}_l},j_{{}_l}$) as classical information to Bob.
Bob applies $2(n-1)$ single-qubit operations at most, namely,
$\bigotimes_{l=1}^{n-1}Z^{i_l}X^{j_l}$, according to the classical
information received from Alice.

For the portion of the shared state in the figure, the initial state
\[
\frac{1}{\sqrt{2}}\sum_{y=0}^1|yy\rangle_{{\rm A}_n{\rm B}_n}
|\kappa(y)\rangle_{{\rm C}_1...{\rm C}_m}
\]
with $|\kappa(y)\rangle_{{\rm C}_1...{\rm C}_m}$ given by Eq.\ (\ref{econkappa})
should be prepared. It is prepared as follows. First we produce
the state
\[
 \frac{1}{\sqrt{2}}\sum_{y=0}^1|yy\rangle_{{\rm A}_n{\rm B}_n}
\otimes \bigotimes_{s=1}^m|{\tilde 0}_s\rangle_{{\rm C}_s}.
\]
This is easily prepared by a single Hadamard gate and a CNOT gate acting
on ${\rm A}_n{\rm B}_n$. The state $|{\tilde 0}_s\rangle_{{\rm C}_s}$
is a basis vector in the $s$th controller's favorite basis. Second, $m$
Hadamard gates are applied to
${\rm C}_1...{\rm C}_m$ individually in their bases. In addition,
the operation
\[
{\rm diag}[1,\exp(i2\pi c_s/p),\exp(i2\pi c_s/p),1]
\]
is applied to ${\rm A}_n{\rm C}_s$ for all $s\in \{1,...,m\}$
in the basis $\{|0\rangle,|1\rangle\}_{{\rm A}_n}
\otimes \{|{\tilde 0}_s\rangle,|{\tilde 1}_s\rangle\}_{{\rm C}_s}$.
The desired initial state for ${\rm A}_n{\rm B}_n{\rm C}_1...{\rm C}_m$
is now achieved. In the teleportation stage,
Alice makes a Bell measurement on ${\rm A'}_n{\rm A}_n$ and
sends Bob the outcome ($i_n,j_n$).
The controllers and Bob follow the protocol as described
in Sec.\ \ref{secEQCQT} or that in Sec.\ \ref{secEBO}. In this process,
Bob can eliminate the phase factors due to controllers' measurements
if $k$ or more controllers agree to allow Bob to obtain the original
state. Finally, Bob applies $Z^{i_n}X^{j_n}$ to ${\rm B}_n$. After all
these steps, he obtains the original state of ${\rm A'}_1...{\rm A'}_n$ in
${\rm B}_1...{\rm B}_n$.

With the above description of the process, we find that
the number of single-qubit operations and that of two-qubit operations
for preparing the initial state of the whole system are $n+m$ for both
operations. The teleportation process involves $n$ Bell
measurements performed by Alice and $m$ single-qubit measurements performed
by controllers. It also involves Bob's recovery operations: (i) at most
$2n$ single-qubit operations corresponding to Alice's measurement
outcomes; (ii) $m$ single qubit operations [on average, ($m-t/2$)
single-qubit operations] corresponding to controllers' measurement
outcomes when the protocol described in Sec.\ \ref{secEQCQT} is employed
[when that in Sec.\ \ref{secEBO} is employed].

In addition, as we have mentioned in Sec.\ \ref{secEBO}, the operations
of (ii) reduce to a single operation in reality for the present setup
while it does not for the setup of Fig.\ \ref{fig2}.

\section{Discussions}\label{sec6}
We have proposed an economical scheme for a $(k,m)$-threshold control of
a quantum teleportation. It uses a shared state of Alice, Bob, and
controllers with the controllers' portion in the state of Eq.\
(\ref{econkappa}), which consists of qubits only. Thus a drawback of a
usual polynomial coding, namely, the required dimension $d\ge p>m-1$ for
each controller's qudit, has been resolved. In addition, it is
straightforward to extend the scheme so that multiple qubits in
Bob's portion are under the threshold control.

Our economical scheme can be seen as a hybrid of a standard
$(m,m)$-threshold controlled quantum teleportation and a
$(k,m)$-threshold classically-controlled quantum teleportation,
as we have mentioned in Sec.\ \ref{secEQCQT}. It should be noted that
this has been realized by a nontrivial protocol using the state
(\ref{econkappa}). The scheme has a good redundancy for the securement of
the $(k,m)$ threshold: (i) To modify the phase factors owing to
controllers' measurements, $k$ or more classical keys are required for Bob.
(ii) To make disagreeing controllers perform measurements, at most
($m-k$) solicits should be sent from Bob.
In fact, the standard controlled teleportation in the context of $(m,m)$
threshold can be used in the context of $(k,m)$ threshold by stating
only (ii) in its protocol. The advantage of our economical scheme over
this simple extension is, thus, the redundant securement.

The scheme is, however, not as economical as a classically-controlled
quantum teleportation, as we have discussed in Sec.\ \ref{secCCQT}.
A quantum threshold-control scheme is certainly more expensive
than a classical threshold-control scheme. It is thus recommended to
assess the trade-off between the benefit and the economicalness to choose
an appropriate scheme.

The benefit to distribute qudits (qubits in our scheme) among
controllers is to make the recovery operation of Bob unfixed unless
controllers make measurements. This makes the protocol robust against
Bob's attack to the keys: In order for cheating, he needs both
physical accesses to at least $k$ controllers' qudits and information
on their measurement bases. One may however claim that careless
controllers tend to lose both of them at once in a real world.
Our economical scheme possesses a clearer advantage also: it can prevent a
violation of a voting schedule, i.e. it can prevent Bob from recovering 
the original state before an official voting time, unless controllers
violate the schedule all together.

There is one drawback in our protocol.
In case we use the scheme of Sec.\ \ref{sec3}, controllers who disagree
to the teleportation do not have to measure their systems. In contrast,
in our economical protocol, a disagreeing controller has to measure her/his
qubit if a solicit is sent from Bob. Our protocol can be broken by a
controller who does not follow this regulation. It seems an important
drawback at a glance, but there is a quick solution: one can easily find
out which controller cheats in the protocol. Any controller who does not
send a measurement outcome despite a solicit sent from noncheating Bob is a
cheater.

A clever cheater, however, may report an opposite measurement outcome
and/or a wrong key instead of being quiet. It has been well-known that
a basic secret sharing is not robust against dishonest participants who
report wrong keys. There have been several proposals to remedy this
drawback in classical secret sharing schemes \cite{TW88,R93,R99,T03}.
These classical schemes are easily combined into our scheme in order to
find cheating in the keys $c_s$, as is clear from the protocol.
Nevertheless, a protocol to find cheaters sending wrong messages as
measurement outcomes should be newly constructed. One way to achieve
this task is to add a supercontroller who grasps controllers' states by
entangling her/his systems and their corresponding systems. Let the
supercontroller know their measurement bases. Then the supercontroller
can check the measurement outcomes afterward. It is hoped that a more
sophisticated way will be developed.

Finally, we discuss a well-known strategy to use controlled quantum
teleportation as a secret sharing to hide a quantum state as a secret
\cite{D05}.
Any ($k,m$)-threshold controlled quantum teleportation in the stage
after Alice's Bell measurements is regarded as a ($k,m$)-threshold
quantum secret sharing: Alice's original state to be recovered in
Bob's side is regarded as a secret and the controllers are regarded
as participants sharing the secret. A drawback of this approach is that
the original state is recovered in the system of Bob's side;
participants who try to cooperate for recovery should gather at this
side or ask a dealer for proxy.

A solution to avoid this drawback
is to construct $m$ initially identical copies of the entire system ${\rm K}$
for the ($k,m$)-threshold controlled quantum teleportation, where
${\rm K}={{\rm A}'}_1{\rm A}_1...{{\rm A}'}_n{\rm A}_n{\rm B}_1...{\rm
B}_n{\rm C}_1...{\rm C}_m$. Let us write a copy as
${\rm K}^u={{\rm A}'}_1^u{\rm A}_1^u...{{\rm A}'}_n^u{\rm A}_n^u
{\rm B}_1^u...{\rm B}_n^u{\rm C}_1^u...{\rm C}_m^u$ with $u\in\{1,...,m\}$.
We send the system ${\rm B}_1^v...{\rm B}_n^v$, together with classical
information obtained by the Bell measurements on each of
${{\rm A}'}_1^v{\rm A}_1^v,...,{{\rm A}'}_n^v{\rm A}_n^v$, to the $v$th
participant ($v\in\{1,...,m\}$). The $v$th participant should also
receive ${\rm C}_{v}^1,...,{\rm C}_{v}^m$, namely the
$v$th control systems of all ${\rm K}^u$'s.
The $v$th participant can recover the original state in the
system ${\rm B}_1^v...{\rm B}_n^v$ when she/he gets to
know the measurement results on $k$ or more among
${\rm C}_{1}^v,...,{\rm C}_{m}^v$,
including her/his own, and corresponding classical keys if they
are used in the scheme.

This approach possesses the following
benefit. The original quantum secret sharing is limited to $m<2k-1$
due to the no-cloning theorem when an unknown quantum state is
a secret \cite{C99}. A controlled quantum teleportation with a
($k,m$)-threshold control is not limited by the no-cloning theorem
because the secret, namely, Alice's original state, is teleported to
Bob's system. The approach is practical when we use classical
keys for controlling a quantum teleportation. The resource required for
this case is $m$ classical keys and $m$ copies of the system that
consists of $n$ EPR pairs and the $n$-qubit original state.
Of course, the $m$ copies of the system are reduced to one copy
in case participants may gather in a particular place or may use a
trusted dealer for proxy.

There seems to be no serious drawback of using classical control
because the classical keys can be securely distributed and a scheduled
vote is not interposed usually for a secret sharing. The robustness of
our economical scheme is effective when an untrusted Bob exists.
This is due to the fact that the measurement basis of each controller
can be hidden. In this sense, our economical scheme might be used for a
secret sharing to build in robustness against physical access
attacks by malicious participants who try to cheat. Nevertheless, it is
not attractive to spend many EPR pairs despite the reduction in the
resource for controllers' portion. A choice of a proper scheme
is dependent on the demand of participants when a controlled quantum
teleportation is applied to a secret sharing.

We have discussed the advantage and disadvantage of our scheme in which
the dimension of each controller's qudit is reduced to two. This
reduction is indeed significant for physical realization of the
threshold control of a quantum teleportation. Nevertheless, it is
controversial as to which extent a controlled quantum teleportation
should be performed with quantum resources. The answer depends on the
application and as to which party is trusted, as is clear from the above
discussions.

\section{Summary}\label{sec7}
We have proposed an economical protocol for $(k,m)$-threshold
controlled quantum teleportation. This protocol uses qubits distributed
to controllers; hence we have achieved the reduction in the dimension of
each qudit from a prime $p>m-1$ to two. In addition,
we have shown an economization in the number of Bob's recovery
operations.

\section*{Acknowledgments}
A.S. and M.N. are supported by ``Open Research Center'' Project
for Private Universities: matching fund subsidy from MEXT.
R.R. is supported by the Grant-in-Aid from JSPS (Grant No. 1907329).
M.N. would like to thank a partial support of the Grant-in-Aid
for Scientific Research from JSPS (Grant No. 19540422).


\begin{thebibliography}{99}
\bibitem{B93} C.~H.~Bennett, G.~Brassard, C.~Cr\'{e}peau, R.~Jozsa,
A.~Peres, and W.~K.~Wootters, ``Teleporting an unknown quantum state via
dual classical and Einstein-Podolsky-Rosen channels'',
Phys. Rev. Lett., vol.70, pp.1895-1899, 1993.
\bibitem{G99} J.~Gruska, Quantum Computing,
McGraw-Hill, London, 1999.
\bibitem{NC2000} M.~A.~Nielsen and I.~L.~Chuang,
Quantum Computation and Quantum Information,
Cambridge University Press, Cambridge, 2000.
\bibitem{Z00}J.~Zhou, G.~Hou, S.~Wu, and Y.~Zhang,
``Controlled Quantum Teleportation'',
LANL arXiv: quant-ph/0006030.
\bibitem{A03} N.~B.~An,
``Teleportation of coherent-state superpositions within a network''
Phys. Rev. A, vol.68, pp.022321-1-6, 2003.
\bibitem{Y04} C.-P.~Yang, S.-I.~Chu, and S. Han,
``Efficient many-party controlled teleportation of multiqubit quantum
information via entanglement'', Phys. Rev. A, vol.70,
pp.022329-1-8, 2004.
\bibitem{D05} F.~G.~Deng, C.-Y.~Li, Y.-S.~Li, H.-Y.~Zhou, and Y.~Wang,
``Symmetric multiparty-controlled teleportation of an arbitrary
two-particle entanglement'', Phys. Rev. A, vol.72, pp.022338-1-8, 2005.
\bibitem{M07} Z.-X.~Man, Y.-J.~Xia, and N.~B.~An,
``Economical and feasible controlled teleportation of an arbitrary
unknown $N$-qubit entangled state'',
J. Phys. B: At. Mol. Opt. Phys., vol.40, pp.1767-1774, 2007.
\bibitem{KM06} D.~Kenigsberg and T.~Mor, ``Secure Controlled
Teleportation'', LANL arXiv: quant-ph/0609028.
\bibitem{B92} C.~H.~Bennett, F.~Bessette, G.~Brassard, L.~Salvail, and
J.~Smolin, ``Experimental quantum cryptography'',
J. Crypt., vol.5, pp.3-28, 1992.
\bibitem{LMC05} H.-K.~Lo, X.~Ma and K.~Chen,
``Decoy state quantum key distribution'',
Phys. Rev. Lett., vol.94, pp.230504-1-4, 2005.
\bibitem{M95} C.~Marand and P.~Townsend,
``Quantum key distribution over distances as long as 30km'',
Opt. Lett., vol.20, pp.1695-1697, 1995.
\bibitem{M96} A.~Muller, H.~Zbinden, and N.~Gisin,
``Quantum cryptography over 23 km of installed under-lake telecom
fibre'', Europhys. Lett., vol.33, pp.335-339, 1996.
\bibitem{H02} T.~Hasegawa, T.~Nishioka, H.~Ishizuka, J.~Abe, K.~Shimizu,
	M.~Matsui and S.~Takeuchi,
``An Experimental Realization of Quantum Cryptosystem'',
IEICE Trans. Fund., vol.E85-A, no.1, pp.149-157, January 2002.
\bibitem{S02} D.~Stucki, N.~Gisin, O.~Guinnard, G.~Ribordy,
and H.~Zbinden, ``Quantum key distribution over 67km with a plug\&play
system'', New J. Phys., vol.4, pp.41-1-8, 2002.
\bibitem{MLK06} I.~Marcikic, A.~Lamas-Linares, and C.~Kurtsiefer,
``Free-space quantum key distribution with entangled photons'',
Appl. Phys. Lett., vol.89, pp.101122-1-3, 2006.
\bibitem{D07} J.~F.~Dynes, Z.~L.~Yuan, A.~W.~Sharpe, and A.~J.~Shields,
``Practical quantum key distribution over 60 hours at an optical fiber
	distance of 20km using weak and vacuum decoy pulses for enhanced
	security'',
Optics Express, vol.15, iss.13, pp.8465-8471, 2007.
\bibitem{BVK98} S.~Bose, V.~Vedral, and P.~L.~Knight,
``Multiparticle generalization of entanglement swapping'',
Phys. Rev. A, vol.57, pp.822-829, 1998.
\bibitem{B97} D.~Bouwmeester, J.-W.~Pan, K.~Mattle, M.~Eibl,
	H.~Weinfurter, and A.~Zeilinger, ``Experimental Quantum
	Teleportation'', Nature, vol.390, no.6660, pp.575-579, 1997.
\bibitem{B98} D.~Boschi, S.~Branca, F.~De~Martini, L.~Hardy, and
S.~Popescu, ``Experimental Realization of Teleporting an Unknown
Pure Quantum State via Dual classical and Einstein-Podolsky-Rosen
channels'', Phys. Rev. Lett., vol.80, no.6, pp.1121-1125, 1998.
\bibitem{M03} I.~Marcikic, H.~de~Riedmatten, W.~Tittel, H.~Zbinden,
and N.~Gisin, ``Long-Distance Teleportation of Qubits at
Telecommunication Wavelengths'', Nature, vol.421, no.6922, pp.509-513, 2003.
\bibitem{B79} G.~R.~Blakley, ``Safeguarding cryptographic keys'',
in Proceedings of AFIPS 1979 National Computer Conference,
New York, 1979, vol.48, pp.313-317, AFIPS Press, Arlington, Va., June 1979.
\bibitem{S79} A.~Shamir, ``How to Share a Secret'',
Commun. ACM, vol.22, no.11, pp.612-613, 1979.
\bibitem{KGH83} E.~D.~Karnin, J.~W.~Greene, and M.~E.~Hellman,
``On Secret Sharing Systems'', IEEE Trans. Inf. Theory, vol.IT-29,
no.1, pp.35-41, January 1983.
\bibitem{K84} S.~C.~Kothari, ``Generalized linear threshold scheme'',
in Proceedings of Crypto'84, Santa Barbara, 1984,
Edited by G.~R.~Blakley and D.~Chaum, Springer-Verlag, New York, 1985,
pp.231-241.
\bibitem{C99} R.~Cleve, D.~Gottesman, and H.-K.~Lo,
``How to Share a Quantum Secret'', Phys. Rev. Lett., vol.83,
pp.648-651, 1999.
\bibitem{Z05} Z.~Zhang, ``Controlled teleportation of an arbitrary
$n$-qubit quantum information using quantum secret sharing of
classical message'', Phys. Lett. A, vol.352, pp.55-58, 2006.
\bibitem{M05} Y.~Murakami, M.~Nakanishi, S.~Yamashita, and K.~Watanabe,
``Cheater Identifiable Quantum Secret Sharing Schemes'',
IPSJ SIG Technical Reports, 2005-CSEC-30(47), pp.337-340, 2005.
\bibitem{I06} S.~Iftene, ``General Secret Sharing Based on the
Chinese Remainder Theorem'', Cryptology ePrint Archive: 2006/166,
http://eprint.iacr.org/2006/166.
\bibitem{KT06} H.~Kunii and M.~Tada, ``A note on information rate for
	fast threshold schemes'', in Proceedings of Computer Security
	Symposium 2006, Kyoto, Japan, October 2006, pp.101-106.
\bibitem{K08} J.~Kurihara, S.~Kiyomoto, K.~Fukushima, and T.~Tanaka,
``A fast (3,n)-threshold secret sharing scheme using exclusive-OR
operations'', IEICE Trans. Fund., vol.E91-A, no.1, pp.127-138, January 2008.
\bibitem{GG03} G.-P.~Guo and G.-C.~Guo,
``Quantum secret sharing without entanglement'',
Phys. Lett. A, vol.310, pp.247-251, 2003.
\bibitem{ZM04} Z.-J.~Zhang and Z.-X.~Man,
``Many-agent controlled teleportation of multi-qubit quantum
information'', Phys. Lett. A, vol.341, pp.55-59, 2005.
\bibitem{MS08} D.~Markham and B.~C.~Sanders,
``Graph states for quantum secret sharing'',
Phys. Rev. A, vol.78, pp.042309-1-17, 2008.
\bibitem{N06} C.~Negrevergne, T.~S.~Mahesh, C.~A.~Ryan, M.~Ditty,
F.~Cyr-Racine, W.~Power, N.~Boulant, T.~Havel, D.~G.~Cory, and
R.~Laflamme, ``Benchmarking Quantum Control Methods on a 12-Qubit System'',
Phys. Rev. Lett., vol.96, pp.170501-1-4, 2006.
\bibitem{LT85} P.~Lancaster and M.~Tismenetsky,
The Theory of Matrices, 2nd Ed., Academic Press, San Diego, 1985.
\bibitem{K68} S.~ Karlin, Total Positivity, Volume I, Stanford
University Press, California, 1968.
\bibitem{A87} T.~Ando, ``Totally Positive Matrices'',
Linear Algebra Appl., vol.90, pp.165-219, 1987.
\bibitem{C98} T.~Craven and G.~Csordas, ``A Sufficient Condition for
Strict Total Positivity of a Matrix'', Linear Multilinear Algebra,
vol.45, pp.19-34, 1998.
\bibitem{PS76} G. P\'{o}lya and G. Szeg\"{o}, Problems and Theorems in
Analysis, Volume II, Springer-Verlag, Berlin, 1976, 4th Ed.
\bibitem{Na06} Proceedings of the symposium ``Quantum Computation: Are
the DiVincenzo Criteria Fulfilled in 2004?'', Edited by M.~Nakahara,
S.~Kanemitsu, M.~M.~Salomaa, and S.~Takagi, published as ``Physical
Realizations of Quantum Computing'', World Scientific, Singapore, 2006.
\bibitem{G02} N.~Gisin, G.~Ribordy, W.~Tittel, and H.~Zbinden,
``Quantum cryptography'', Rev. Mod. Phys., vol.74, no.1, pp145-195, 2002.
\bibitem{TW88} M.~Tompa and H.~Woll, ``How to share a secret with cheaters'',
J. Crypt., vol.1,  pp.133-138, 1989.
\bibitem{R93} J.~Rif\`{a}-Coma, ``How to Avoid the Cheaters Succeeding
in the Key Sharing Scheme'', Designs, Codes, and Crypt., vol.3,
pp.221-228, 1993.
\bibitem{R99} R.~S.~Rees, D.~R.~Stinson, R.~Wei, and G.~H.~J.~van~Rees,
``An application of covering designs: determining the maximum consistent
set of shares in a threshold scheme'', Ars Combin. vol.53,
pp.225-247, 1999.
\bibitem{T03} R.~Tso, Y.~Miao, and E.~Okamoto, ``A new algorithm for
searching a consistent set of shares in a threshold scheme with
cheaters'', in Proceedings of the 6th Information Security and Cryptology
Conference, Edited by J.~I.~Lim and D.~H.~Lee,
Lecture Notes in Computer Science, vol.2971, pp.377-385,
Springer-Verlag, Berlin, 2003.
\end{thebibliography}
\end{document}